\input{aipcheck}

\documentclass[
    ,final            
  ]
  {aipproc}

\layoutstyle{6x9}

\newcommand{\cN}{\mathcal N}
\newcommand{\bea}{\begin{eqnarray}}
\newcommand{\eea}{\end{eqnarray}}
\def\ra{\rangle}
\def\la{\langle}
\def\s{\sigma}
\def\m{\mu}
\def\corr#1{\left\langle #1 \right\rangle}
\newcommand{\cO}{\mathcal O}
\def\a{\alpha}
\def\b{\beta}
\newcommand{\nn}{\nonumber}
\def\Dim{\textrm{Dim}}
\def\vmn{ V^{\otimes m } \otimes \bar V^{\otimes n } }
 
\def\cO{\mathcal{O}} 
\def\bX{{\bf{X}}} 

\begin{document}

\title{   Schur-Weyl duality as an instrument    of  Gauge-String duality            }


\author{Sanjaye Ramgoolam }{
  address={ Centre for Research in String Theory, Department of Physics,\\ 
Queen Mary, University of London,\\  Mile End Road, London E1 4NS, UK
}
}

\begin{abstract}
A class of mathematical dualities 
have played a central role in mapping states in gauge theory 
to states in the spacetime string theory dual. This includes the classical 
Schur-Weyl duality between symmetric groups and Unitary groups, 
as well as generalisations involving Brauer and Hecke algebras.
 The physical string dualities involved  include examples from 
the  AdS/CFT correspondence as well as the string dual 
of two-dimensional Yang Mills. 

\end{abstract}

\maketitle


\section{ Introduction  }
The AdS/CFT correspondence \cite{malda,gkp,wit} 
between string theory on $ AdS_5 \times S^5 $
and $ N=4 $ SYM with $U(N)$ gauge group  in four dimensions has provided 
a remarkably rich setting where gravitons as well as 
classical  brane and spacetime 
geometries can be seen as  emergent phenomena from local operators 
in the quantum gauge theory.   Early work showed how
single-particle  Kaluza-Klein gravitons  in spacetime correspond to 
gauge invariant single trace operators in the dual gauge theory. 
The computation of correlation functions in  gauge theory and on the  
gravity side  gave evidence for this map \cite{agmoo}.
Some qualitatively new features of finite $N$
were highlighted as the stringy exclusion principle 
\cite{malstrom,jevram,hrt}. A geometrical realisation of the 
stringy exclusion principle was discovered  in the properties  of 
super-symmetric rotating branes or giant gravitons \cite{mst}.  
These $S^3$-shaped branes  in the $S^5$ of spacetime,  
grow in size as the angular momentum $L$ 
is increased. This increase in size was related \cite{mst} to the 
growth in transverse size of stringy objects due to 
the stringy uncertainty principle  (see \cite{yoneya} for a review).  When the 
angular momentum  $L$ reaches $N$, the size of the worldvolume $S^3$ 
reaches the size of the background $ S^5$.  
The angular momentum cannot grow any further and the classical
SUSY brane solution ceases to exist. 
Other solutions were found \cite{giantgravitons}  
which are also $S^3$-shaped but live in the $AdS_5$ and
carry the same angular momentum as the $S^5$ or sphere-giants. 
These new giant graviton solutions are called dual giants
or AdS-giants. These AdS-giants have no upper bound on the 
size of the $S^3$-worldvolume and no upper bound on the angular momentum.   

An attempt to find local gauge invariant duals of the giant gravitons 
was initiated in \cite{bbns}, where sub-determinant operators 
were proposed as  gauge theory duals of single sphere giants.
This lead to the classification of half-BPS multi-trace 
operators and a more complete map to giant gravitons in \cite{cjr}. 
The gauge invariant operators were described in terms of a 
Young diagram basis, which was shown to have diagonal two-point 
functions. This basis lead to a proposal of multi-column Young
 diagrams as multi-sphere giants, and multi-row Young diagrams
as multi-AdS giants. Three-point functions were
 computed in terms of group-theoretic
Littelwood-Richardson coefficients. Subsequently the 
map of giant gravitons to Young diagrams was given elegant confirmation 
by constructing open strings attached to multiple-giants
(see \cite{bbfh,rob} and refs. therein). 
The Young diagram picture  of half-BPS operators
constructed from a complex matrix  lead to 
a system of $N$ free fermions in a harmonic oscillator  \cite{cjr,ber,taktsu}. 
Recent work on the classification of half-BPS supergravity
solutions discovered the  free fermion structure in spacetime, 
in the guise of  black-white colourings of a plane. These colourings  
provide boundary conditions for non-singular solutions in spacetime
and reflect the occupied and unoccupied regions of  
free fermion phase space \cite{LLM}. Strings in certain 
BMN limits of LLM backgrounds 
have also been studied from the gauge theory point of view \cite{ccs}.

This talk is based mainly
on \cite{bhr,kimram}, where two generalisations of \cite{cjr} 
are considered. In \cite{bhr} we consider eighth-BPS 
operators at zero coupling, with a view to finding the gauge invariant 
operators dual to eighth-BPS giant gravitons. In \cite{kimram} 
we consider a non-supersymmetric sector with a view to understanding 
systems of branes and anti-branes. 
More complete lists of relevant references are
given in these papers. In this talk I would like to highlight 
an important mathematical equivalence, Schur-Weyl duality,  
as a recurrent principle in the map between gauge theory 
states and stringy spacetime states  \cite{cjr,bhr,kimram}. 
SW duality had previously played an instrumental role 
in the string theory of two dimensional Yang Mills
\cite{gt,cmr}.   

\section{ Half BPS : One complex matrix   }

Half-BPS operators, which have scaling dimension $ \Delta $  equal to 
$U(1)$ charge $J$, are constructed from a complex matrix $ X = \Phi_1 +
 i \Phi_2 $. 
$\Phi_1,\Phi_2$ are two of the six hermitian matrices, transforming in the 
adjoint of the $U(N)$ gauge group of $ \cN =4 $ supersymmetric 
gauge theory. The matrix $ X  $ defines a linear map from 
an $N$-dimensional vector space $ V$ to $V$.  
It is useful to consider $ { \bf X  } =
  X  \otimes X  \cdots \otimes X  $ 
as a  map from $ V^{\otimes n } $ to   $ V^{\otimes n } $. 
Multi-trace gauge invariant operators can be constructed 
by composing $ { \bf X } $ with permutations. Take a simple 
case  of $n=2$. Choose a basis $ e_{ i } $ for $V$. 
\bea 
tr_{ V^{\otimes 2 } }  ( X \otimes X ) & \equiv &  tr_2 ( X \otimes X )  \cr 
& = &  \la   e^{i_1} \otimes e^{ i_2} |  ( X \otimes X ) 
 | e_{i_1} \otimes e_{ i_2} \ra \cr 
& = &  X^{ i_1}_{i_1} X^{i_2}_{i_2 }  = ( tr X )  (tr X )   
\eea 
Now consider  $tr_{ V^{\otimes 2 } }   ( ( X \otimes X ) \sigma ) $. 
The permutation $ \sigma $ is an element of the symmetric group 
of permutations of $2$ elements and  acts by permuting the two factors 
of $ V \otimes V$. Explicitly,  
\bea 
&& \sigma | e_{i_1} \otimes e_{ i_2} \ra = | e_{i_2} \otimes e_{ i_1 } \ra 
\eea 
The trace of $   ( X \otimes X ) \sigma  $ is  
\bea 
 tr_{ 2 }      ( ( X \otimes X ) \sigma ) &=&  
                  \la e^{i_1} \otimes e^{ i_2}  |   ( X \otimes X ) \sigma 
                | e_{i_1} \otimes e_{ i_2} \ra \cr 
&=&  \la e^{ i_1 } \otimes e^{i_2} | X \otimes X | e_{i_2} \otimes e_{ i_1 } \ra \cr 
&=&  X^{i_1}_{i_2} X^{i_2}_{i_1} = tr ( X^2 ) 
\eea

In general  $ tr_n (  { \bf X} \sigma ) $     depends on the 
cycle structure of the permutation $ \sigma $. 
Any permutation can be decomposed into a product of cyclic permutations, 
e.g the permutation which takes $ (1,2,3,4,5) $ to $ ( 3, 1,2,  5,4 ) $ 
is the product of cycles $ ( 123) (45)$.   Let $C_i ( \sigma ) $ be the 
number of cycles of length $i$ in $\sigma $. Then   
\bea  
tr_n  ( \sigma \bX  ) = X^{i_1 }_{i_{\sigma(1) } } \cdots ~ 
 X^{i_n }_{i_{\sigma(n) } } 
 = \prod_{ i=1 }^{n}    ( tr X^i )^{C_i ( \sigma ) }  
\eea 
In the gauge theory $ X$ transforms in  the adjoint of $U(N)$. 
This adjoint action comes from  the transformation of $V$ 
as the fundamental of $U(N)$. 
\bea 
&& U | e_i > = U^j_{i}  | e_j \ra  
\eea  
The action on $ V^{ \otimes n } $ is given by 
\bea 
U ~  | e_{ i_1} \otimes e_{i_2} \cdots \otimes e_{ i_n } \ra 
= U^{j_1}_{ i_1} \cdots U^{j_n}_{i_n }  | e_{ j_1} \otimes e_{j_2} 
\cdots \otimes e_{ j_n } \ra
\eea 
It is an easy exercise to check that the actions of $U(N)$ and $ S_n$ 
on $ V^{ \otimes n } $ commute with each other. We can also easily 
check that the action of the Lie algebra  $u(N)$ generated 
by the elementary matrices $E_{ij}$ commutes with $S_n $.   
In fact the complete
set of operators in  $ V^{ \otimes n } $ commuting with  $U(N)$
is given by the action of the group algebra of $S_n$. Conversely the 
commutant of $ S_n$ is the enveloping algebra of $u(N)$.  
This has a consequence, via the double commutant theorem,  also called the 
double centraliser theorem (see e.g  \cite{ramthesis}, \cite{goodwall}),  
that the decomposition  of  $ V^{ \otimes n } $
in terms of $ U(N) \times S_n $ is very simple
\bea\label{schwey} 
V^{\otimes n } = \oplus_{ R  } V_R^{ U(N) }  \otimes V_R^{ S_n } 
\eea 
Here $R$ runs over Young diagrams with first column of length 
no greater than $N$. The vector space $ V_R^{ U(N) }$ is the irreducible 
representation  (irrep.) of $U(N)$ corresponding to Young diagram $R$
and $   V_R^{ S_n } $ is the irrep. of $S_n $
corresponding to Young diagram $ R$.

Several powerful consequences follow from (\ref{schwey}), when it is 
combined with standard group theory properties, such as orthogonality of 
characters. These are used along with 
  the free field correlators 
\bea 
 \la X^i_j  ( X^{\dagger} )^k_l \ra = \delta^i_l \delta^k_j  
\eea 
 to show \cite{cjr} that the Schur-basis of matrix operators 
\bea 
\chi_{ R } ( X ) = { 1 \over n ! } 
\sum_{ \s \in S_n } \chi_R ( \s  ) tr_{n} ( \s { \bf X } ) 
\eea  
satisfies the orthognality 
\bea 
\la \chi_{ R } ( X )  \chi_{ S } ( X^{\dagger}  ) \ra 
 = \delta_{ R S } f_R 
\eea 
The normalisation factor $f_R $ is 
\bea\label{normfac}  
f_R = { n! Dim R \over d_R }  
\eea 
where $ Dim R $ is the dimension of  $ V_R^{ U(N) }$ 
 and $d_R $ is the  dimension of  $   V_R^{ S_n } $.
We have dropped the trivial spacetime dependence of these correlators.  
Local operators in CFT correspond to states in CFT. 
Duality maps states in the CFT to states in spacetime. 
The column lengths of the Young diagrams map to angular momenta 
of sphere giants. The row lengths map to AdS giants. 
These descriptions are valid in different regimes, where 
the semiclassical descriptions in terms of S and AdS  giants 
respectively are appropriate. The fact that the S-giants have angular 
momentum cutoff at $N$ whereas the AdS-giants have unbounded 
angular momentum is naturally explained by the Young diagram description
\cite{cjr}. This is compatible with the backreacted geometries 
of LLM  and with the description of open strings attached to 
multi-giants. The normalisation factor (\ref{normfac}) encodes 
the stringy exclusion principle simply in that it vanishes 
for Young diagrams with column lengths greater than $N$. 
Three-point functions are 
\bea\label{threepoint}  
\la \chi_{R_1 } ( X ) \chi_{R_2} ( X ) \chi_{R_3} (X^{\dagger} ) 
= g ( R_1 , R_2 ; R_3 ) f_{R_3} 
\eea 
where $g ( R_1 , R_2 ; R_3 ) $ is the Littlewood-Richardson 
coefficient for the coupling of Young diagrams $R_1,R_2 $ with $n_1,n_2$ boxes 
into  Young diagram $R_3$ of $n_3=n_1+n_2$ boxes.

For the application of Schur-Weyl duality (\ref{schwey})
in the string theory of 
two dimensional $U(N)$  Yang Mills see section 4 of \cite{cmr}.
In that context cycle structures of permutations correspond to winding
numbers  of multi-string states, whereas free fermions 
are related to the irreducible representation basis 
of observables in the $ U(N)$ gauge theory. The counting 
of string worldsheet maps to spacetime, described according to 
classic theorems on  branched covers in mathematics, by 
appropriate sums over permutations, is recovered from the
large $N$ expansion of the gauge theory using the fundamental 
equation (\ref{schwey}).  In \cite{cr} 
the projectors for the different Young diagrams 
in (\ref{schwey}),  along with simple diagrammatic techniques 
for manipulations in tensor spaces (for any rank $n$), 
are used to  study the factorisation equations of CFT in the  
half-BPS context. The ref. \cite{copto} contains a review 
of the underlying operator-state correspondence of CFT 
and applications of the factorisation equations to 
the probabilistic interpretation of correlators. 
Useful mathematical references on Schur-Weyl duality 
are \cite{zel,fulhar}.  
Schur-Weyl duality is an application of the  double-commutant theorem. 
For a review of this and several useful related theorems 
see for example section 1 of \cite{halverson} and \cite{ramthesis,goodwall}.

\section{ Eighth-BPS }
The eighth-BPS operators at zero coupling are constructed from 
holomorphic combinations of three complex matrices $ X_1 , X_2 , X_3 $. 
These are charged respectively $ (1,0,0) , ( 0,1,0) , (0,0, 1 )  $
under the $U(1)\times U(1) \times U(1)  $ subgroup of 
$ SO(6)$ global symmetry group of $ \cN =4 $ SYM theory. 
The holomorphic subspace is preserved by the $U(3)$ subgroup of 
$SO(6)$.  The general holomorphic covariant operators are 
\bea 
X_{a_1~ j_1}^{i_1} ~ X_{a_2 ~ j_2 }^{i_2}  ~ \cdots ~  X_{a_n ~ j_n }^{i_n }  
\eea 
The indices  $a_1 .. a_n $ each take values from $1$ to $3$
in the fundamental of $U(3)$. The discussion generalizes 
to $ U(M)$ where $a_i $ are in the fundamental of $U(M)$.  
The $i,j$ indices take values from $1$ to $N$. 
The covariant operators can be written as 
\bea\label{genopcov} 
X_{a_1} \otimes X_{ a_2}\otimes  \cdots \otimes X_{ a_n } 
\eea 
and viewed as maps from $ V^{ \otimes n } $ to $ V^{\otimes n } $.  
The operators in (\ref{genopcov}) transform as 
the $V_M^{\otimes n } $ of $U(M)$ where $V_M $ 
is the fundamental $M$-dimensional representation of 
$U(M)$.

 In solving the problem of finding a diagonal basis of gauge
 invariant operators constructed from (\ref{genopcov}) 
 we encounter two roles for Schur-Weyl duality. 
 As in the half-BPS case of section 2, we use 
\bea\label{schwey1}  
  V^{ \otimes n } =  \oplus_{ R } V_R^{ U(N) }  \otimes V_R^{ S_n }  
\eea 
$R$ runs over Young diagrams with no more than $N$ rows. 
 Since we also want to organise the operators 
 according to their $U(M)$ transformation properties 
 we also have 
\bea\label{schwey2}  
    V_M^{ \otimes n }  = \oplus_{ \Lambda  } V_{ \Lambda } ^{ U(M) }
  \otimes V_{ \Lambda } ^{ S_n }  
\eea 
$\Lambda$ runs over Young diagrams with no more than $M$ rows.

\subsection{ The main result  } 

Employing the decompositions (\ref{schwey1}) (\ref{schwey2}) 
the  gauge invariant operators 
in a diagonal basis are labelled by $R$,  a Young diagram with $n$ boxes 
associated to the $U(N)$ gauge symmetry ; by $\Lambda$ another
 Young diagram with $n$ boxes associated to the  
$U(M) $ global  symmetry, along with  some additional labels explained below.  
 Consider the class of operators 
 with $ \mu_1 $ copies of $X_1$, $ \mu_2 $ copies of 
 $X_2 $, $\mu_3 $ copies of $X_3 $. Clearly at zero coupling, 
 using the free field correlator 
\begin{equation}\label{basiccorrelatorUN}
  \corr{ (X_{a})^i_j (X_{b}^\dagger)^k_l} = \delta_{ab}\, \delta^i_l \delta^k_j
\end{equation}
 operators characterised by different choices of 
 $ \mu_1 , \mu_2  , \mu_3 $ are orthogonal to each other. 
 Hence the diagonalisation problem in the eighth-BPS 
 sector can be solved for each fixed $  \mu = ( \mu_1 , \mu_2 , \mu_3 )  $
 with $ \mu_1 + \mu_2 + \mu_3 = n  $. 
 Let 
\bea 
{\bf X}^{  \m }  = X_{1}^{\otimes \m_1} \otimes \cdots \otimes
X_{M }^{\otimes \m_M} 
\eea 
The following operators provide a diagonal basis for two-point functions, 
hence for the metric on operators defined using two-point functions 
\bea\label{theops} 
   {\cO}^{ \Lambda \mu , R }_{ \b, \tau }= \frac{1}{n!}\sum_{\a } B_{j \b } \; S^{\tau  ,}{}^{  \Lambda }_{  j
  }\;{}^{R}_{p }\;{}^{R}_{q}\;\; D_{pq}^R(\a) ~~  tr (\a\;
  {\bf X}^{\mu} )
\nn\label{theops} 
\eea 
The label $ \beta $ runs over the number of times
the trivial representation of $H_{  \mu}   \equiv S_{\mu_1} \times S_{\mu_2} \times  \cdots  \times S_{\mu_M} $ 
is contained in $ S_{ n } $, where $n= \mu_1 + \cdots + \mu_M $.   The label $ \tau $ runs 
over the number of times $\Lambda $ appears in the $S_n$ Clebsch-Gordan 
decomposition of $R \otimes R$.   $D_{pq}^R(\a)$ are the 
matrix elements of the permutation $ \alpha $ in an orthogonal basis. 
$ S^{\tau  ,}{}^{  \Lambda }_{  j
  }\;{}^{R}_{p }\;{}^{R}_{q}$ is a Clebsch-Gordan coefficient 
 for the coupling of $ R \otimes R $ to $ \Lambda $. 
The $B$-factor $ B_{j \b } $ is a { \it branching coefficient}, giving 
a change of basis in the
subspace of the irrep. $ \Lambda $ of $S_n$ 
invariant under the subgroup $ H_{  \mu}$. The counting of the operators 
is easily written down in terms of Clebsch-Gordan multiplicities 
$ C ( R , R , \Lambda )$ and Littlewood-Richardson multiplicities 
for coupling horizontal Young diagrams of lengths 
$ \mu_1, \mu_2 \cdots \mu_M $ into the Young diagram $\Lambda $.
This agrees with counting derived  by expanding 
partition functions using properties of characters \cite{dolan}.

The $ {\cO}^{ \Lambda \mu , R }_{ \b, \tau } $ can be proved to have  orthogonal 
2-point functions \cite{bhr} 
\begin{displaymath} 
  \langle  {\cO}^{ \Lambda_1 \mu^{(1)} ,R_1}_{ \beta_1,\tau_1 } {\cO}^\dagger{}^{\Lambda_2 \mu^{(2)}  ,R_2}_{ \beta_2 ,
  \tau_2 } \rangle  
\end{displaymath}
\begin{displaymath}
= \delta^{ \m^{(1)} \m^{(2)} }
\delta^{{\Lambda}_1 {\Lambda}_2 }  \delta_{ \beta_1 \beta_2 }\delta^{R_1 R_2 } \delta_{ \tau_1 \tau_2 }
  \frac{ |H_{\m^{(1)}}|  \Dim R_1}{d_{R_1}^2}\label{orthoresult} 
\nn 
\end{displaymath} 
In this way, the problem of computing a diagonal basis 
for the multi-matrix operators has been reduced to the 
computation of standard $S_n$ group theory quantities 
which are available for example in 
\cite{hamermesh}. Likewise 3-point (and higher-point)
 functions can be computed 
in terms of this type of group theory data (\cite{bhr}).

\subsection{ Weak coupling and the chiral ring } 

The above basis (\ref{theops}) solves the problem of diagonalising 
the two-point functions at zero coupling. In order to compare to eighth-BPS 
giant gravitons, we need to consider the strong coupling
spectrum. The spectrum of eighth-BPS operators 
at weak coupling can be obtained by finding the subspace 
of the zero coupling operators which are orthogonal to 
operators such as $ tr (  [X_1,X_2] [ X_1, X_2 ]  ) $ 
which are descendants. The zero-coupling diagonalisation 
allows us to write some neat formulae for the 
inverse metric in the trace basis. This gives some 
useful information \cite{bhr} on the problem of finding a 
diagonal basis of eighth-BPS operators at weak coupling, 
but it remains an  open problem. Further work 
will be interesting as it will allow comparison with 
strong coupling discussions \cite{berencr,bglm}.

\section{ The $ X  , X^{\dagger} $ sector  : Branes and anti-branes }

The operators  $ \chi_R ( X ) $ of section 2 correspond to states 
with $ \Delta  = J $ (scaling dimension equals angular momentum)
 and are giant gravitons. 
The   $ \chi_R ( X^{\dagger}  ) $ 
have $ L_0 = - J $. They correspond to branes moving in the 
opposite direction. The time reversed brane solutions are 
solutions of the anti-brane action. Hence  $ \chi_R ( X^{\dagger}  ) $
correspond to anti-giant gravitons. Operators constructed from 
both $ X , X^{\dagger} $ correspond to composite systems
built from branes and anti-branes. The consideration of diagonal bases
for these composite operators can be done using a generalisation 
of Schur Weyl duality involving Brauer algebras \cite{kimram}, 
for the case where $U(N)$ is acting on tensor products of 
the fundamental and anti-fundamental, i.e 
$ V^{\otimes m } \otimes {\bar V}^{\otimes n } $

We have seen that $ S_n  $ is the commutant  of 
$U(N)$ (or $GL(N)$)  acting on $ V^{\otimes n } $
leading to (\ref{schwey}). 
$S_m \times S_n$ is contained in  commutant  
of $U(N)$ ( or $GL(N)$ ) 
 acting in  $ V^{\otimes  m } \otimes \bar V^{\otimes  n } $. 
But we also need contractions, which along with 
the permutations, generate the Brauer algebra $ B_N ( m  , n ) $. 
Hence Schur-Weyl duality, in this case (see for example \cite{halverson}),
   states that 
\bea\label{nonchirschwe}  
V^{  \otimes m } \otimes \bar V^{ \otimes n }
= \oplus_{ \gamma } V_{ \gamma }^{   U(N) } \otimes 
V_{\gamma } ^{ B_N (m,n)   } 
\eea 
It gives the decomposition of the tensor product $\vmn$ 
in terms of irreps of $U(N)$ and $B_N ( m,n)$.   
$ \gamma $ runs over  sets of integers 
$ ( \gamma_1 , \gamma_2 , \cdots , \gamma_N )$ obeying
$ \gamma_1 \ge \gamma_2 \ge \cdots \ge \gamma_N $.
The set of positive integers defines $ \gamma_+$ which is a 
partition of $ m - k $  while the 
negative integers define a partition $ \gamma_{-} $ of 
$ n - k$. Here $k$ is an integer lying between $0$ and $min ( m,n)$. 
 Equivalently $ \gamma_+ $  determines a Young diagram 
with $m-k$ boxes, $\gamma_- $ one of $n-k$ boxes.   A choice of  $ \gamma $ 
 is equivalent to a choice of $ ( k , \gamma_+ , \gamma_- ) $. 
If we write $\gamma_+ $ as a Young diagram, with row lengths 
equal to the parts in the partition, $c_1 ( \gamma_+ ) $ is defined 
as the length of the first column.   It follows from the above 
definitions  that $ c_1 ( \gamma_+ ) + c_1 ( \gamma_- ) \le N $.

The Brauer algebra contains a class of elements,
\bea\label{QOPS}  
Q^{\gamma}_{A , i  j }
\eea
 called symmetric branching
operators in \cite{kimram}, which have the properties 
\bea\label{branchmatmult}  
Q^{\gamma_1}_{A , i  j } Q^{\gamma_2}_{B , k  l }
= \delta_{\gamma_1 \gamma_2} \delta_{A B } \delta_{j k }
 Q^{\gamma_1}_{A , i  l  }
\eea
and 
\bea 
 h Q^{\gamma }_{A , i  j }  h^{-1} =  Q^{\gamma }_{A , i  j }
\eea 
for $ h \in S_m \times S_n $.  The index $A$ consists of a 
pair of labels $ ( \alpha , \beta ) $ for irreps. 
of $S_{m} \times S_{n} $. The indices $i,j$ each 
run over the multiplicity of the irrep. $A$ in the $S_m \times S_n$ 
decomposition of the irrep. $ \gamma $ of $ B_N(m,n)$. 

The operators
 $ tr_{m,n}  \bigl ( \Sigma ( Q^{\gamma }_{A, i j}  \bigr )   
( \bX \otimes  \bX^{\dagger} )  \bigr )$ 
 diagonalise the two-point functions. 
\bea\label{brauerdiagcor}  
&& \la   tr_{m,n}   
\bigl ( \Sigma ( Q^{\gamma_2 ~ \dagger }_{A_2 , i_2 j_2} )   
 ( \bX \otimes  \bX^{\dagger} )  \bigr )
  tr_{m,n}  \bigl ( \Sigma ( Q^{\gamma_1 }_{A_1 , i_1 j_1} )   
( \bX \otimes  \bX^{\dagger} )  \bigr )\ra  \cr 
&& = ~~ m! n! \delta_{\gamma_1 \gamma_2 } \delta_{A_1 A_2} 
\delta_{i_1 i_2  } \delta_{j_1 j_2 }~ d_{A_1} ~  \Dim ~  \gamma_1 
\eea 
The trace is in $ V^{  \otimes m } \otimes \bar V^{ \otimes n }$, 
$ \bX $ denotes the $m$-fold tensor product 
$ X \otimes \cdots \otimes X $, and  $ \bX^{ \dagger}  $ is 
the $n$-fold tensor product 
$   X^{\dagger}  \otimes \cdots \otimes X^{\dagger}  $. 
$ \Sigma $ is a map from $ B_N ( m,n ) $ to $S_{m+n} $ defined 
in \cite{kimram}. $\Dim ~  \gamma_1  $ 
is the dimension of the $U(N)$ irrep. labelled by 
$\gamma_1 $.

A special class of operators are those corresponding to $ k=0$. 
In these cases, $ \alpha = \gamma_+ , \beta = \gamma_- $, 
the reduction from Brauer to $ S_m \times S_n$ gives a unique irrep 
with multiplicity $1$.  In this $k=0$ case the 
$Q$-operators become projection operators, denoted 
as $P_{ R \bar S } $ (using the relabelling  $ \alpha = R , \beta = S $
to match the notation of two dimensional Yang Mills \cite{gt}) 
which are related 
to the coupled characters $ \chi_{ R \bar S } ( U  ) $ 
 in two dimensional Yang Mills. 
The operators 
$tr_{ m,n} ( \Sigma ( P_{ R \bar S } ( \bX \otimes \bX^{\dagger} )) )$  
 are proposed as ground states of a composite system 
made from a giant graviton corresponding to Young diagram 
$R$ and an anti-giant corresponding to Young diagram 
$S$. Higher $k$ operators are interpreted as excited states of 
brane-anti-branes.  The stringy exclusion principle for individual 
giant gravitons imposes the condition $ c_1(R ) \le N $
and $ c_1 ( S ) \le N $, which says that Young diagrams 
for $U(N)$ cannot have more than $N$ boxes in the first column. 
The surprising lesson from the Brauer algebra description 
of brane-anti-branes and their excited states is that 
there is no ground state corresponding to a giant-anti-giant 
if the condition $ c_1(R ) + c_1( S ) \le  N $ is violated.   

The map $ \Sigma $ appearing in the formula for the operators in 
(\ref{brauerdiagcor}) is also used in the explicit construction 
of Brauer algebra projectors in \cite{kimram}. As a corollary of 
that construction, there is a new formula for the coupled 
dimension 
\bea\label{newcoupform} 
 { d_R^2  d_S^2 
\over  { Dim R \bar S } } =    { m!^2 n!^2 \over { ( m+n )!^2 }} 
\sum_{ T }    { d_T^2  \over Dim T }  g(R, S ; T ) 
\eea 
where $R,  S$ are Young diagrams with  $m,n$ boxes respectively, 
$T$ has $ m+n$ boxes  and $g(R,S; T)$ are Littlewood-Richardson coefficients. 
This formula is developed further to derive a new formulation 
of the nonchiral large $N$ expansion of 2d Yang Mills \cite{kimram2}.

\section{ Summary and Outlook } 

Schur-Weyl duality has provided a powerful set of tools 
for the mapping of gauge theory states to 
strings or branes in spacetime or large deformations of spacetime. 
The simplest and most thoroughly understood cases 
are those  of half-BPS operators in $\cN = 4 $ super Yang Mills 
theory in four dimensions.  Recent work has been undertaken 
on the quarter and eighth BPS sectors \cite{bhr},
 as well as a non-supersymmetric sector \cite{kimram}, 
providing bases of multi-matrix operators which diagonalise 
the free field two-point functions in the field theory.
Clebsch-Gordan coefficients of symmetric groups and Branching 
coefficients of Brauer algebras have appeared in the solutions 
of these diagonalisation problems. 

Some other recent work related to this theme is described here.  
The computation of the one-loop dilatation operator in 
the $SU(2)$ sector \cite{Beisert} using the basis of 
\cite{bhr} was undertaken \cite{tom}. 
A closely related diagonalisation  method, inspired 
by earlier work on the construction of strings attached to 
branes, has been proposed \cite{bcr}. The Brauer algebras
used in the diagonalisation problem of the $ X , X^{ \dagger} $ 
sector have  been used to simplify the complete large $N$ expansion of 
two dimensional Yang Mills, giving  it a 
new  holomorphic string interpretation \cite{kimram2}. 
The application of Schur-Weyl duality  in the case of $q$-deformed 
two dimensional Yang Mills  was done in \cite{drt} where 
Hecke algebras,which are   $q$-deformations of symmetric group algebras
and  Schur-Weyl duals of $q$-deformed $U(N)$, provide 
$q$-deformations of branched cover counting problems. 
The extension of work on diagonalising correlators in four dimensions from 
$U(N)$ to $SU(N)$ has produced interesting results 
(section 10 of \cite{cr} and \cite{robgwyn,tomsu}).  

Interesting open questions include the understanding of 
finite $N$ cutoffs (stringy exclusion principle) 
on gauge invariant operators   in less supersymmetric contexts, 
and in non-supersymmetric contexts in terms of spacetime physics
of branes and black holes.  
We would like all the cutoffs on matrix operators in the gauge theory 
to have simple spacetime stringy  interpretations like 
the ones for sphere giant gravitons.  It has been suggested recently 
that the stringy exclusion principle is related to black hole complementarity
\cite{iipol}. It is an interesting problem to develop this idea 
using detailed information about the counting and construction of 
operators in gauge theory at finite $N$. Since the diagonal bases 
naturally encode the finite $N$ cutoffs in the normalisations 
(e.g. eq. (\ref{normfac})) of  two-point functions 
(even at zero coupling) they should be useful tools
in this direction. A clear interpretation  of three-point functions 
such as (\ref{threepoint}) in terms of giant graviton moduli spaces 
in spacetime would also  be very desirable, perhaps 
using known connections between Littlewood-Richardson coefficients 
and the cohomology of Grassmanians \cite{fulton}.

Schur-Weyl duality has  been
a suprisingly effective technical tool in gauge-string duality,
capturing crucial aspects of the map 
between gauge theory states and spacetime string   
theory states, both for two dimensional and four dimensional gauge theory. 
It is undoubtedly going to continue to play this role 
and provide valuable information on many interesting physical questions on 
 gauge theory, especially  in relation to  its stringy spacetime dual. 
It is  natural to wonder if an appropriately 
enriched version of Schur-Weyl duality might actually give 
a complete mathematical expression of  the background independent 
content of gauge string duality.

\begin{theacknowledgments}
 I would like to  thank my collaborators 
 Tom Brown, Stefan Cordes, 
 Steve Corley, Sebastian de Haro, Robert de Mello Koch,
  Pei Ming Ho,  Greg Moore, Antal Jevicki,
 Paul Heslop,  Yusuke Kimura, Mihail Mihailescu, Radu Tatar, 
Alessandro Torrielli,
 Nick Toumbas for   enjoyable collaborations on projects related to the
 theme of this talk over the years, and  Hally Ingram,  
  Rodolfo Russo  and Gabriele Travaglini for useful conversations.  
 I was delighted to attend  the ``Ten years of AdS/CFT'' conference
 and present this talk.  I  gratefully acknowledge  the support of an STFC Advanced Fellowship.  
\end{theacknowledgments}

\end{document}